\documentclass[useAMS,usenatbib]{mn2e}
\usepackage{ulem}
\usepackage{graphicx}
\usepackage{amsmath}
\usepackage{amssymb}
\usepackage{color}
\usepackage{subfig}
\usepackage[breaklinks,colorlinks,citecolor=blue,linkcolor=red]{hyperref} 
\usepackage[all]{hypcap}

\newcommand\msun{\, \rm M_\odot}
\newcommand\rsun{\, \rm R_\odot}

\newcommand\yr{{\, \rm yr}}

\newcommand\eout{{e_{\rm out}}}
\newcommand\ein{{e_{\rm in}}}

\newcommand\aout{{a_{\rm out}}}
\newcommand\ain{{a_{\rm in}}}
\newcommand\msmbh{{M_{\rm SMBH}}}

\title[Massive binary star mergers in galactic nuclei]{Massive binary star mergers in galactic nuclei: implications for blue stragglers, binary S-stars and gravitational waves}
\author[Fragione and Antonini]{{Giacomo Fragione$^{1}$\thanks{E-mail: giacomo.fragione@mail.huji.ac.il}, Fabio Antonini$^{2}$}\\
$^1$Racah Institute for Physics, The Hebrew University, Jerusalem 91904, Israel
\\
$^2$Faculty of Engineering and Physical Sciences, University of Surrey, Guildford, Surrey, GU2 7XH,
United Kingdom
}

\begin{document}

\maketitle

\begin{abstract}
Galactic nuclei are often found to contain young stellar populations and, in most cases, a   central supermassive black hole (SMBH).
Most known massive stars are found in binaries or higher-multiplicity systems, and in a galactic nucleus the gravitational interaction with the SMBH can affect their 
long-term evolution. In this paper, we study the orbital evolution of stellar binaries
near SMBHs using high precision $N$-body simulations, and including tidal forces and Post-Newtonian corrections to the motion. We focus on the Lidov-Kozai (LK) effect induced by the SMBH on massive star binaries. We investigate how the properties of the merging binaries change with varying the SMBH mass, the slope of the initial mass function, the distributions of the binary orbital parameters and the efficiency in energy dissipation in dissipative tides. We find that the fraction of merging massive binary stars is in the range $\sim 4\%$--$15\%$ regardless of the details of the initial distributions of masses and orbital elements. For a Milky Way-like nucleus, we find a typical rate of binary mergers $\Gamma\approx 1.4\times10^{-7} \yr^{-1}$. The merger products of massive binaries can be rejuvenated blue-straggler stars, more massive than each of their original progenitors, and G2-like objects. Binary systems that survive the LK cycles can be source of X-rays and gravitational waves, observable with present and upcoming instruments.
\end{abstract}


\section{Introduction}

The proximity of the Galactic Centre (GC) offers the unique opportunity to study star, compact objects and gas dynamics in the vicinity of a supermassive black hole (SMBH). The big advance in present (and upcoming) instruments has made possible to resolve physical scales on the order of a fraction of a parsec, and thus to investigate in detail the physical and dynamical properties of individual stars \citep{scho2014,scho2018}. These precise observations serve as reference to test theoretical predictions for a variety of phenomena that are believed to take place not only in our GC, but also in other galactic nuclei. These mechanisms include  the formation of a stellar cusp \citep{alex09,perets2007,chen2015,bar16,baum2018,frasari18}, the ejection of hypervelocity stars \citep{hills1988,yut2003,ses2007,sari10,brw14,frcapkr17}, the formation and evolution of S-stars \citep{gou2003,peretal2009,gill2017}, the origin of tidal disruption events \citep{rees1988,chen2009,stone2016,fralei18}, and the gravitational wave emission from merging black holes \citep{antper2012,aha2016,bartos2017,fragrish2018,frlei2018,gondan2018,hamer18,2017ApJ...846..146P,hoa18,arc2019}.

Most of the stars are believed to be born in binaries, triples or higher multiplicity systems \citep{rag10,tok14a,tok14b,rid15}. Our Galactic Centre contains a large population of young massive O-type stars, many of which have been observed to reside in a stellar disk and probably were born \textit{in-situ} as a consequence of the fragmentation of a gaseous disk formed from an infalling gaseous clump \citep{genz10}. Three binaries have been observed in the GC within $\sim 0.1$ pc, namely an equal-mass $50\msun$ binary with a period of $19.5$ days \citep{ott1999}, a binary with a period of $224$ days and an eccentricity of $0.3$, and an eclipsing Wolf-Rayet binary with a period of $2.3$ days \citep{pfu2014}. Recently, \citet{jia2019} have found that the S-star S0-27 could be the first potential astrometric binary candidate in the GC. The implications of the presence of a binary population in the GC have been under scrutiny only recently. \citet{alexpf14} showed that long-period binaries can be used to constrain the cusp of dark remnants in the GC. \citet{li2017} calculated that, if the massive star S0-2 is a binary, the eclipse probability would be $\sim 30$\%-$50$\%. \citet{naoz2018} discussed the role of stellar binaries in biasing the properties of the disk of young stars observed in the GC. Nevertheless, depicting self-consistently the formation and evolution of binary (or multiple) stars in the crowded environment of the GC is not straightforward. The only comprehensive (even though limited) study of binary dynamics over a relaxation timescale dates back to \citet{hop09}, who claimed that the binary fraction can be as low as $\sim 10\%$ at the SMBH influence radius ($\sim 2$ pc) in our GC. 

\citet{2010ApJ...713...90A} and
\citet{2011ApJ...731..128A} were the first to argue that
any binary star within the SMBH influence radius form a three-body system with the SMBH as the third distant companion; as a consequence, the binary undergoes inclination and eccentricity oscillations due to the Lidov-Kozai (LK) mechanism if its orbit is sufficiently inclined with respect to the SMBH.
If the excursion in eccentricity is significant and dissipative processes such as tidal friction are highly efficient during periapsis passages, the binary shrinks its orbit and can merge \citep[see also][for a review]{naoz2016}. \citet{prod2015} studied the properties of the binary mergers due to LK oscillations and found that most bianries merge within one LK cycle \citep{ghez2005,gill2012,gill2013}. More recently, \citet{steph2016M} coupled the secular equation of motions to post-main-sequence stellar evolution of binaries that become dynamically decoupled from the gravitational perturbations of the SMBH. 
 
Previous calculations adopted a secular approximation of the equation of motion to study the triples made up of an SMBH and a stellar binary \citep[with the exception of][]{2010ApJ...713...90A}. However, the secular theory could not be anymore an adequate description of the three-body dynamics whenever the stellar binary undergoes rapid oscillations in the angular momentum and eccentricity \citep{2014ApJ...781...45A,antognini14,grish18}. For these cases, direct precise $N$-body simulations, including regularization schemes, are required to follow accurately the orbits of the objects up to the possible final merger.

In this paper, we revisit the SMBH-induced mergers of stellar binaries orbiting in its vicinity by means of $N$-body simulations, including tidal forces and Post-Newtonian (PN) terms. In particular, we focus on massive binaries which would lead to the formation of compact objects (COs), possibly sources of gravitational waves (GWs). We evolve a three-body system consisting of an inner stellar binary (of masses $m_1$ and $m_2$, semi-major axis $\ain$ and eccentricity $\ein$), and an outer binary comprised of the SMBH (of mass $\msmbh$) and the centre of mass of the inner binary (semi-major axis $\aout$ and eccentricity $\eout$). We study how the properties of the merging binaries change with varying the SMBH mass, the slope of the initial mass function (IMF), the distribution of the inner and outer orbital parameters, and the efficiency in energy dissipation due to dissipative tides. Finally, we discuss the observational consequences of the mechanism we study in this paper. 

The paper is organized as follows. In Section~\ref{sect:timescales}, we discuss the relevant timescales in galactic nuclei. In Section~\ref{sect:binprop}, we discuss the properties of the binary stars under exam, while, in Section~\ref{sect:nbody}, we present the results of our numerical simulations. Finally, in Section~\ref{sect:disc}, we discuss the implications of our findings and, in Section~\ref{sect:conc}, we draw our conclusions.

\section{Timescales in galactic nuclei}
\label{sect:timescales}

In the dense stellar environment of galactic nuclei, several dynamical processes can take place and affect the evolution of the stellar binary population.

Stars and COs move in the smooth near-Keplerian potential of the SMBH, which dominates the dynamics within the radius of influence $R_h$, beyond which the potential of the SMBH is overcome by the galactic field \citep{mer13}. On timescales much longer than the orbital period, the microscopic fluctuations of the potential make stars' energy and angular momentum diffuse, as a result of continuous non-coherent scattering events with other stars and COs. This process happens over a 2-body relaxation timescale \citep{1987degc.book.....S}
\begin{eqnarray}
T_{2B}&=&1.6\times 10^{10}\mathrm{yr}\ \left(\frac{\sigma}{300\ \mathrm{km s}^{-1}}\right)^3\left(\frac{m}{\msun}\right)^{-1}\times\nonumber\\
&\times&\left(\frac{\rho}{2.1\times 10^{6}\msun\ \mathrm{pc}^{-3}}\right)^{-1}\left(\frac{\ln \Lambda}{15}\right)^{-1} \ .
\label{eqn:t2b}
\end{eqnarray}
Here, $\rho$ and $\sigma$ are the 1-D density and velocity dispersion in the galactic nucleus, respectively, $\ln \Lambda$ is the Coulomb logarithm and $m$ is the average stellar mass. The classical result by \citet{bahcall76} shows that a population of equal-mass objects forms a power-law density cusp around an SMBH, $n(r)\propto r^{-\alpha}$, where $\alpha=7/4$. For multi-mass distributions, lighter and heavier objects develop shallower and steeper cusps, respectively \citep{Fre06,ale09,aharon16,baumg18}, while source terms (as star formation and binary disruptions) tend to make the cusp steeper as well \citep{aharon15,frasari18}. Recent observations of the Milky Way's centre showed that the slope of the cusp appears to be shallower \citep[$\alpha\sim 5/4$;][]{gall18,scho18}. Only galactic nuclei harbouring SMBHs less massive than $\sim 10^7\ \mathrm{M}_{\odot}$ \footnote{We use the $\msmbh$-$\sigma$ relation \citep[see e.g.][]{mer01}.} have typical evolutionary timescales ($T_{2B}$) small enough to make the effects of the uncorrelated stellar interactions important within a Hubble time
\citep{2009ApJ...694..959M}. Therefore a relaxed steady-state cusp is expected to have been built up only for $\msmbh \lesssim 10^7\ \mathrm{M}_{\odot}$.

\begin{table*}
\caption{Models: SMBH mass ($M_\mathrm{SMBH}$), slope of the initial mass function ($\beta$), mass-ratio distribution of massive stars ($f(q)$), semi-major axis distribution ($f(\ain)$), $\ein$ distribution, slope of the outer semi-major axis distribution ($\alpha$), $\eout$ distribution, time-lag factor ($\tau$), merger fraction ($f_{\rm merge}$).}
\centering
\begin{tabular}{ccccccccc}
\hline
$M_\mathrm{SMBH}$ (M$_\odot$) & $\beta$ & $f(q)$ & $f(\ain)$ & $f(\ein)$ & $\alpha$ & $f(\eout$) & $\tau$ (s) & $f_{\rm merge}$ \\
\hline\hline
$4\times 10^6\msun$ & $2.3$ & -         & Kroupa 1995     & thermal & $1.5$ & thermal & $0.66$  & $0.12$ \\
$1\times 10^7\msun$ & $2.3$ & -         & Kroupa 1995     & thermal & $1.5$ & thermal & $0.66$  & $0.15$ \\
$1\times 10^8\msun$ & $2.3$ & -         & Kroupa 1995     & thermal & $1.5$ & thermal & $0.66$  & $0.13$ \\
$4\times 10^6\msun$ & $1.7$ & -         & Kroupa 1995     & thermal & $1.5$ & thermal & $0.66$  & $0.13$ \\
$4\times 10^6\msun$ & $2.3$ & uniform   & Kroupa 1995     & thermal & $1.5$ & thermal & $0.66$  & $0.11$ \\
$4\times 10^6\msun$ & $2.3$ & $m_1=m_2$ & Kroupa 1995     & thermal & $1.5$ & thermal & $0.66$  & $0.13$ \\
$4\times 10^6\msun$ & $2.3$ & -         & log-uniform     & thermal & $1.5$ & thermal & $0.66$  & $0.05$ \\
$4\times 10^6\msun$ & $2.3$ & -         & uniform         & thermal & $1.5$ & thermal & $0.66$  & $0.15$ \\
$4\times 10^6\msun$ & $2.3$ & -         & Duquennoy+ 1991 & thermal & $1.5$ & thermal & $0.66$  & $0.11$ \\
$4\times 10^6\msun$ & $2.3$ & -         & Sana+ 2012      & thermal & $1.5$ & thermal & $0.66$  & $0.04$ \\
$4\times 10^6\msun$ & $2.3$ & -         & Kroupa 1995     & thermal & $2.0$ & thermal & $0.66$  & $0.11$ \\
$4\times 10^6\msun$ & $2.3$ & -         & Kroupa 1995     & thermal & $2.5$ & thermal & $0.66$  & $0.10$ \\
$4\times 10^6\msun$ & $2.3$ & -         & Kroupa 1995     & thermal & $3.0$ & thermal & $0.66$  & $0.11$ \\
$4\times 10^6\msun$ & $2.3$ & -         & Kroupa 1995     & uniform & $1.5$ & thermal & $0.66$  & $0.10$ \\
$4\times 10^6\msun$ & $2.3$ & -         & Kroupa 1995     & thermal & $1.5$ & uniform & $0.66$  & $0.11$ \\
$4\times 10^6\msun$ & $2.3$ & -         & Kroupa 1995     & thermal & $1.5$ & thermal & $0.066$ & $0.14$ \\
$4\times 10^6\msun$ & $2.3$ & -         & Kroupa 1995     & thermal & $1.5$ & thermal & $6.6$   & $0.11$ \\
\hline
\end{tabular}
\label{tab:models}
\end{table*}

On timescales smaller than $T_{\rm 2B}$, but longer than the stars orbital period, the residual torque due 
to the $\sqrt{N}$ 
deviation of the stellar potential from spherical symmetry 
becomes relevant in affecting the  orbits of stars \citep{rauch96}. Both the direction and magnitude of the angular momentum (hence eccentricity) diffuse typically over a resonant relaxation timescale \citep{kocs15}
\begin{equation}
T_{\rm RR}=9.2\times 10^{8}\ \mathrm{yr}\ \left(\frac{\msmbh}{4\times 10^6 \msun}\right)^{1/2}\left(\frac{a_{out}}{0.1\ \mathrm{pc}}\right)^{3/2}\left(\frac{m}{\msun}\right)^{-1}\ .
\label{eqn:trr}
\end{equation}

On even shorter timescales, vector resonant relaxation changes the direction (hence the relative inclination) of the outer orbit angular momentum on a timescale \citep{kocs11,kocs15}
\begin{eqnarray}
T_{\rm VRR}&=&7.6\times 10^{6}\ \mathrm{yr}\ \left(\frac{\msmbh}{4\times 10^6 \msun}\right)^{1/2}\times\nonumber\\
&\times & \left(\frac{a_{\rm out}}{0.1\ \mathrm{pc}}\right)^{3/2}\left(\frac{m}{\msun}\right)^{-1}\left(\frac{N}{6000}\right)^{-1/2}\ ,
\label{eqn:tvrr}
\end{eqnarray}
where $N$ is the number of stars within $a_{\rm out}$.

In the context of LK oscillations, vector resonant relaxation plays a role, since it may affect the initial inclination of the inner and outer orbit of the binaries on timescales comparable to or even shorter than the Lidov-Kozai timescale \citep{lid62,koz62}
\begin{equation}
T_{\rm LK}=\frac{8}{15\pi}\frac{m_{\rm tot}}{M_{\rm SMBH}}\frac{P_{\rm out}^2}{P_{\rm bin}}\left(1-e_{\rm out}^2\right)^{3/2}\ .
\label{eqn:tlk}
\end{equation}
In the previous equation, $P_{\rm bin}$ and $P_{\rm out}$ are the binary period and its orbital period with respect to the SMBH, respectively, and $m_{\rm tot}=m_1+m_2+\msmbh$. Whenever the initial mutual orbital inclination of the inner and outer orbit $i_0\sim 40^\circ$-$140^\circ$, the eccentricity of the inner binary reaches (quadruple order of approximation)
\begin{equation}
e_{\rm in}^{\rm max}=\sqrt{1-\frac{5}{3}\cos^2 i_0}\ ,
\label{eqn:emax}
\end{equation}
starting from a circular orbit.
As $i_0$ approaches $\sim 90^\circ$, the inner binary eccentricity approaches almost unity, and the binary can merge. In the case the outer orbit is eccentric, the inner eccentricity can reach almost unity even if the initial inclination is outside of the $i_0\sim 40^\circ$-$140^\circ$ Kozai-Lidov range \citep[octupole order of approximation;][]{naoz13a,li14}. This happens over the octupole timescale
\begin{equation}
T_{\rm oct}=\frac{1}{\epsilon}T_{\rm LK}\ ,
\label{eqn:tlkoct}
\end{equation}
where the octupole parameter is defined as 
\begin{equation}\label{oc1}
\epsilon={m_1-m_2\over m_1+m_2}\frac{\ain}{\aout}\frac{\eout}{1-e_{\rm out}^2}\ .
\end{equation}

Nevertheless, LK cycles can be suppressed by additional sources of precession, such as non-dissipative tides, that operate on a timescale \citep{kisel1998,egg2001}
\begin{eqnarray}
T_{\rm Tide}&=&\frac{8\ain^{13/2}}{15[G(m_{\rm 1}+m_{\rm 2})]^{3/2}}\frac{(1-\ein^2)^5}{8+12\ein^2+\ein^4}\nonumber\\
&\times& \left[2\frac{m_2}{m_1}k_1 R_1+2\frac{m_1}{m_2}k_2 R_2\right]^{-1}\ ,
\label{eqn:ttidal}
\end{eqnarray}
where $k_1$, $R_1$ and $k_2$, $R_2$ are the apsidal motion constant and radius of the two stars in the binary, respectively, or general relativistic precession, that operates on a typical timescale \citep{1964PhRv..136.1224P}
\begin{equation}
T_{\rm GR}=\frac{a_{\rm in}^{5/2}c^2(1-e_{\rm in}^2)}{3G^{3/2}(m_{\rm 1}+m_{\rm 2})^{3/2}}\ .
\end{equation}

Finally, binaries may evaporate due to dynamical interactions with field stars in the dense environment of a galactic nucleus when 
\begin{equation}
\frac{E_b}{(m_1+m_2)\sigma^2}\lesssim 1\ ,
\end{equation}
where $E_b$ is the binary internal orbital energy. This happens on an evaporation timescale \citep{binntrem87}
\begin{eqnarray}
T_{\rm EV}&=&3.2\times 10^{7}\ \mathrm{yr} \left(\frac{m_1+m_2}{2\msun}\right)\left(\frac{\sigma}{300\ \mathrm{km s}^{-1}}\right)\left(\frac{\msun}{m}\right)\nonumber\\
&\times&\left(\frac{a_{\rm in}}{1\ \mathrm{AU}}\right)^{-1}\left(\frac{2.1\times 10^{6}\msun\ \mathrm{pc}^{-3}}{\rho}\right)\left(\frac{15}{\ln \Lambda}\right)\ .
\label{eqn:binevap}
\end{eqnarray}

\section{Binary properties}
\label{sect:binprop}

In our models, we focus on massive binaries, which would lead to the formation of compact objects COs. We simply assume that every star in the mass range $8 \msun$--$20\msun$ will form a neutron star (NS), while even more massive stars collapse to a black hole (BH). We sample the masses of the stars from an initial mass function
\begin{equation}
\frac{dN}{dm} \propto m^{-\beta}\ ,
\label{eqn:bhmassfunc}
\end{equation}
in the mass range $8\msun$-$150\msun$. We adopt in our models $\beta=2.3$ (canonical \citet{kroupa2001} mass function\footnote{In this mass range, this is also the classical \citet{salp1955} mass function.}). We run an additional model where we take $\beta=1.7$ for binaries originating from stellar disks, which are observed to be
relatively more massive and follow a top-heavy initial mass function \citep{lu2013}. We compute stellar radii from \citep{dem91}
\begin{equation}
R_*=
\begin{cases}
1.06\ (m/\msun)^{0.945}\rsun& \text{$ m< 1.66\msun$},\\
1.33\ (m/\msun)^{0.555}\rsun& \text{$ m> 1.66\msun$}.
\end{cases}
\end{equation}
Following \citet{frlgk2018}, we assume that the main-sequence (MS) lifetime $\tau_{\rm MS}$($m$) provides a good approximation for the progenitor lifetime whenever the progenitor mass $m \le 18$ M$_{\odot}$, justified by the fact that the MS lifetimes of low-mass stars greatly exceed that of every other evolutionary phase, typically by several orders of magnitude \citep[e.g.][]{clayton68,iben91,maeder09}. We are ignoring any metallicity dependence in the MS lifetime, since it would only weakly affect it. For the MS lifetime, we assume \citep{hansen94}
\begin{equation}
\tau_{\rm MS}(m) = \tau_{\rm 0}\Big( \frac{m}{\msun} \Big)^{-2.5}
\label{eqn:mstime}
\end{equation}
with $\tau_{\rm 0} = 10^{10}$ yr. For progenitor masses $m > 18$ M$_{\odot}$, we impose a fixed total lifetime of $7$ Myr, roughly in agreement with stellar evolution models, which predict a near-constant lifetime for massive stars at low metallicity \citep[e.g.][]{iben91,hurley00,maeder09}. Thus, our final estimate for the total progenitor lifetime is
\begin{equation}
\tau_{\rm p} = \max(\tau_{\rm MS}(m), 7\,{\rm Myr}).
\end{equation}

In our fiducial model, we assume that the masses of the two stars in the binary are independent from each other, sample them randomly from the above mass function and randomly pair them. However, observations show the mass-ratio distribution of massive binaries to be rather uniform \citep{sana2012,kob2014}. We add a model with a uniform mass-ratio distribution, and also an additional model where the stars in the binary are equal mass ($m_1=m_2$).

For what concerns the semi-major axis of our binaries, we adopt a \citet{kroupa1995} period distribution function 
\begin{equation}
f(\log_{10} P)\propto \frac{\log_{10} P-1}{45+(\log_{10} P-1)^2}\ .
\end{equation}
where P is the orbital period in units of day.
To study how the results depend on the assumed distribution of periods (or semi-major axes), we run a model with a log-uniform distribution of semi-major axes and an additional model where we sample from a uniform distribution of semi-major axes. We also add a model where the binary orbital periods are drawn from \citet{duq1991}
\begin{equation}
f(\log_{10} P)\propto \exp\left({-\frac{\log_{10} P-\mu_P}{2\sigma_P^2}}\right)\ ,
\end{equation}
where $\mu_P (d)=4.8$ and $\sigma_P (d)=2.8$. Finally, we study an additional set of simulations considering the intrinsic period distribution of O stars derived by \citet{sana2012}
\begin{equation}
f(\log_{10} P)\propto (\log_{10} P)^{-0.55}\ .
\end{equation}
In all the models, we sample the inner semi-major axes in the range $0.1$-$50$ AU \citep{sana2012}.

The inner and outer eccentricity are drawn from a thermal distribution, but we also consider models where they are sampled from an uniform distribution.

As discussed in the previous section, stars and COs tend to form a cusp within the SMBH sphere of influence. In the present study, we assume that the distribution of the binary number densities follow a cusp with slope $\alpha=1.5$. To study how the cusp slope affects the distribution of mergers, we consider different values of the slope of the density distribution $\alpha=2$--$2.5$--$3.0$. For the maximum outer semi-major axis, we fix it to $a_{\rm out}^{M}=0.1(\msmbh/4\times 10^6\msun)^{2/9}$ pc \citep{steph2016M,hoa18}. 

The initial mutual inclination $i_0$ between the inner and outer orbit is sampled from an isotropic distribution (i.e. uniform in $\cos i$)\footnote{We note that the distribution of the mutual inclination of the orbital planes could be different from the isotropic distribution, in particular in the in-situ formation scenario via fragmentation of a gaseous disc, where there could be a correlation between the relative inclinations of the orbital planes. In this case, the merger rate may be suppressed and the octupole corrections may become important.}. The other relevant angles are drawn randomly.

\section{N-Body Simulations}
\label{sect:nbody}

We consider three different SMBH masses, i.e. $\msmbh=4\times 10^6\msun$ for a Milky-Way-like nucleus, $\msmbh=10^7\msun$, and $\msmbh=10^8\msun$ for a M31-like nucleus. We summarise the main parameters of our models in Table~\ref{tab:models}. We adopt as \textit{fiducial model} the first model listed in Table~\ref{tab:models}. After we draw the main parameters of the three-body system, we require that the massive binary does not cross the Roche limit of the SMBH at its orbital pericentre distance
\begin{equation}
\frac{\aout}{\ain}> \eta \frac{1+\ein}{(1-\eout)}\left(\frac{3\msmbh}{m_1+m_2}\right)^{1/3}\,.
\label{eqn:hills}
\end{equation}
Following \citet{antper2012}, we set $\eta=4$, since at shorter distances the binary can be unstable. We then integrate the triple SMBH-binary differential equations of motion using the \textsc{archain} code \citep{mik06,mik08}. This code is fully regularized and is able to model the evolution of objects of arbitrary mass ratios and eccentricities with extreme accuracy, even over long periods of time. We include PN corrections up to order PN2.5, and tidal force prescriptions as in \citet{anthl2016}. In particular, the tidal perturbation force on the object of mass $m_j$ due to an object of mass $m_i$, at a distance $r_{ij}$, is set equal to \citep{hut1981}
\begin{equation}
{\bf{F}}_\mathrm{Tide}^{j,i}=-\frac{Gm_i m_j}{r_{ij}^2}\left[3\frac{m_i}{m_j}\left(\frac{R_j}{r_{ij}}\right)^5k_j\left(1+3\frac{\dot{r}_{ij}}{r_{ij}}\tau_j \right){\bf{\hat{r}}_{ij}}\right]\ ,
\end{equation}
where $\tau_j$ and $k_j$ are the time-lag factor (lag of the instantaneous direction of $m_i$ due to the tidal bulge of $m_j$) and apsidal motion constant (dimensionless measure of the distortability of $m_j$), respectively, of the object of mass $m_j$, and $R_j$ its radius. The time-lag factor is $\tau_j\approx 1/nQ_{\rm tide,j}$, where $Q_{\rm tide,j}$ is the tidal dissipation factor, defined as the ratio between the energy stored in the tidal bulge and the energy dissipated per orbit, and $n$ the mean motion. Thus, larger values $\tau$ imply smaller values of $Q$, hence a more efficient dissipation through tidal forces. This corresponds to an orbital decay over a typical friction timescale
\begin{equation}
T_\mathrm{Fric,j}=\frac{t_\mathrm{V,j}}{9(1+2k_j)^2}\left[\frac{\ain(1-e_{\mathrm{in}}^2)}{R_j}\right]^8\frac{m_j^2}{(m_i+m_j)m_i}\ ,
\label{eqn:tfric}
\end{equation}
where $a$ and $e$ are the semi-major axis and eccentricity of the binary orbit, respectively, and $t_\mathrm{V,j}$ is the viscous timescale
\begin{equation}
t_\mathrm{V,j}=\frac{3}{4}\frac{n(1+2k_j)^2}{k_j}\frac{Q_\mathrm{tide,j} R_j^3}{Gm_j}=\frac{3}{4}\frac{(1+2k_j)^2}{k_j}\frac{R_j^3}{Gm_j\tau_j}\ .
\end{equation}
For instance, using Eq.~\ref{eqn:tfric}, we find that $T_{\rm fric}\sim 5\times 10^{14} (0.66\ \mathrm{s}/\tau)(\ain(1-e_{\rm in}^2)/\mathrm{AU})^8$ yr for a massive binary comprised of two stars of $10\msun$ each. In our models, we set the apsidal motion constant to $k_1=k_2=0.014$ and the time-lag factor of dissipative tides to $\tau_1=\tau_2=0.66$ s for both stars in the binary \citep{prod2015}. We add two additional models where we study the possible effect of varying the time-lag factor, which is currently not well constrained, by setting $\tau_1=\tau_2=0.066$ s and $\tau_1=\tau_2=6.6$ s.

\begin{figure} 
\centering
\includegraphics[scale=0.55]{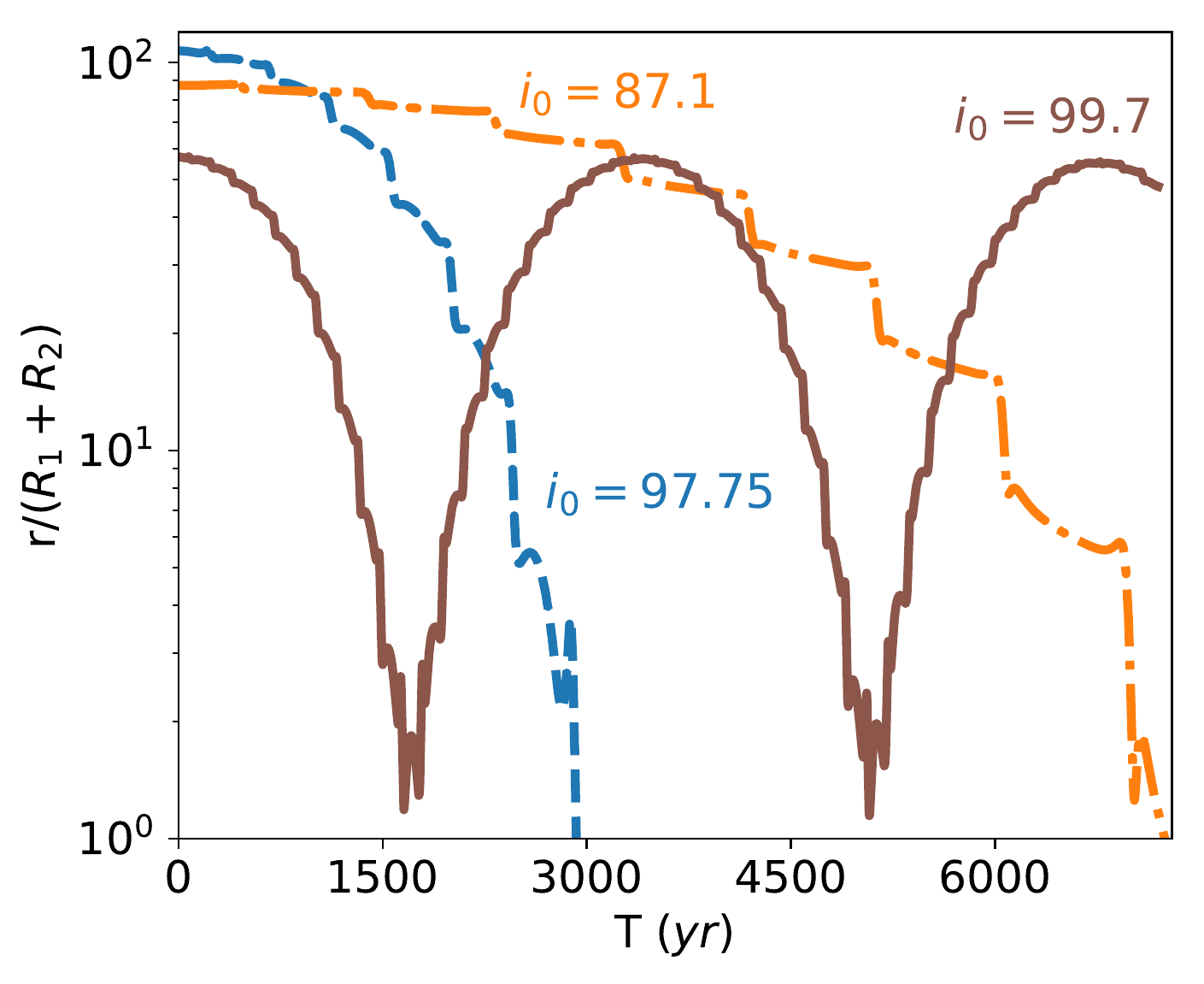}
\caption{Example of three N-body runs: relative distance ($r$) between the stars of the binary in units of the sum of the stars radii ($R_1+R_2$). When $r/(R_1+R_2)<1$, the systems are considered to have merged. Orange dot-dashed line: $m_1=0.96\msun$, $m_2=0.65\msun$, $i_0=87.1^\circ$; brown solid line: $m_1=1.16\msun$, $m_2=3.82\msun$, $i_0=99.7^\circ$; blue dashed line: $m_1=0.97\msun$, $m_2=2.55\msun$ $i_0=97.75^\circ$. The initial semi-major axes and inclinations are $\ain=4.5$ AU, $\ein=0.8$ (orange line), $\ain=2.5$ AU, $\ein=0.6$ (brown line) and $\ain=5.7$ AU, $\ein=0.7$ (blue line), respectively.}
\label{fig:example}
\end{figure}

For each of the $17$ different models, we run $\sim 500$--$1000$ realizations, for a total of $\sim 13000$ simulations. We run simulations up to a total integration time set by the minimum of the MS lifetimes of the two stars in the binary (Eq.~\ref{eqn:mstime}). We stop the integration when we identify a binary merger, namely when
\begin{equation}
\frac{r}{R_1+R_2}<1\ ,
\end{equation}
where $r$ is the relative distance between the two stars, or whenever when one of the two stars overflows its Roche lobe \citep{egg83}. Finally, we note that our binaries typically do not evaporate within their MS lifetime in our region of interest.

Figure~\ref{fig:example} shows the results of three different three-body integrations. As a consequence of the LK mechanism, the binaries with initial inclination $i_0=87.1^\circ$ and $i_0=97.75^\circ$ merge. Typically, most of the binaries merger within one or a few LK cycles \citep{2010ApJ...713...90A,prod2015}. In the case of the binary with initial inclination $i_0=99.7^\circ$, we find that the binary does not merge and its semi-major axis gradually reduces as a consequence of the dissipative tides acting at each pericenter passage. We note that binaries may enter the non-secular regime during the maximum of an LK oscillation and the secular equation of motions are a poor description of the triple dynamics. In this regime, the binary can be driven to a merger before general relativistic effects and tidal precession suppress the LK oscillations.

We now describe the key physical properties of the massive binaries that merge in our models.

\subsection{Inclination distribution}

Most of the massive binaries are expected to be significantly perturbed by the tidal field of the SMBH whenever the binary orbit around the SMBH is sufficiently inclined, nominally in the LK window $i_0\sim 40^\circ$-$140^\circ$, with respect to the inner orbital plane \citep{lid62,koz62}. At the qusruple order of approximation, the binary eccentricity can become almost unity when $i_0\sim 90^\circ$, according to Eq.~\ref{eqn:emax}.

\begin{figure} 
\centering
\includegraphics[scale=0.55]{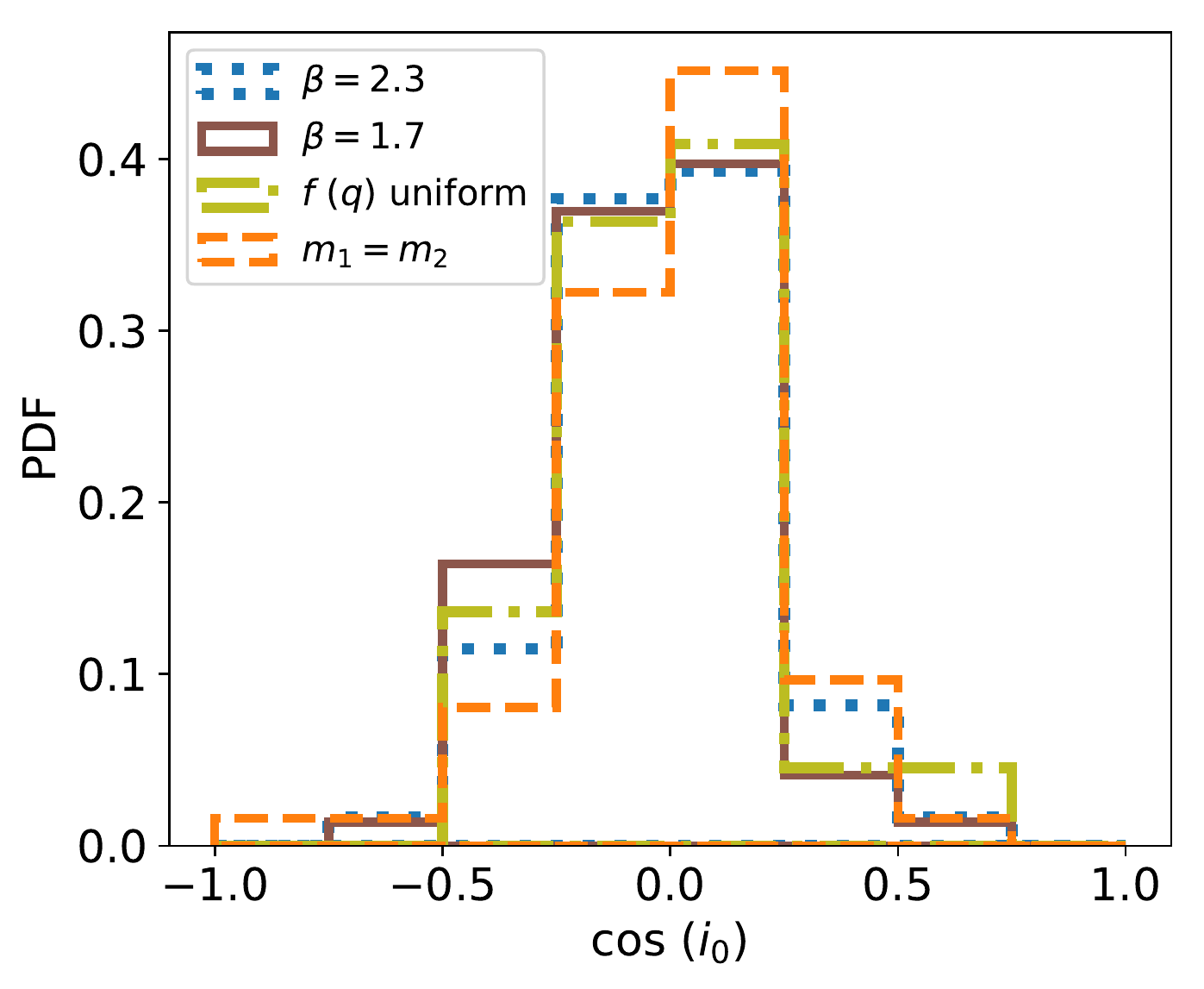}
\includegraphics[scale=0.55]{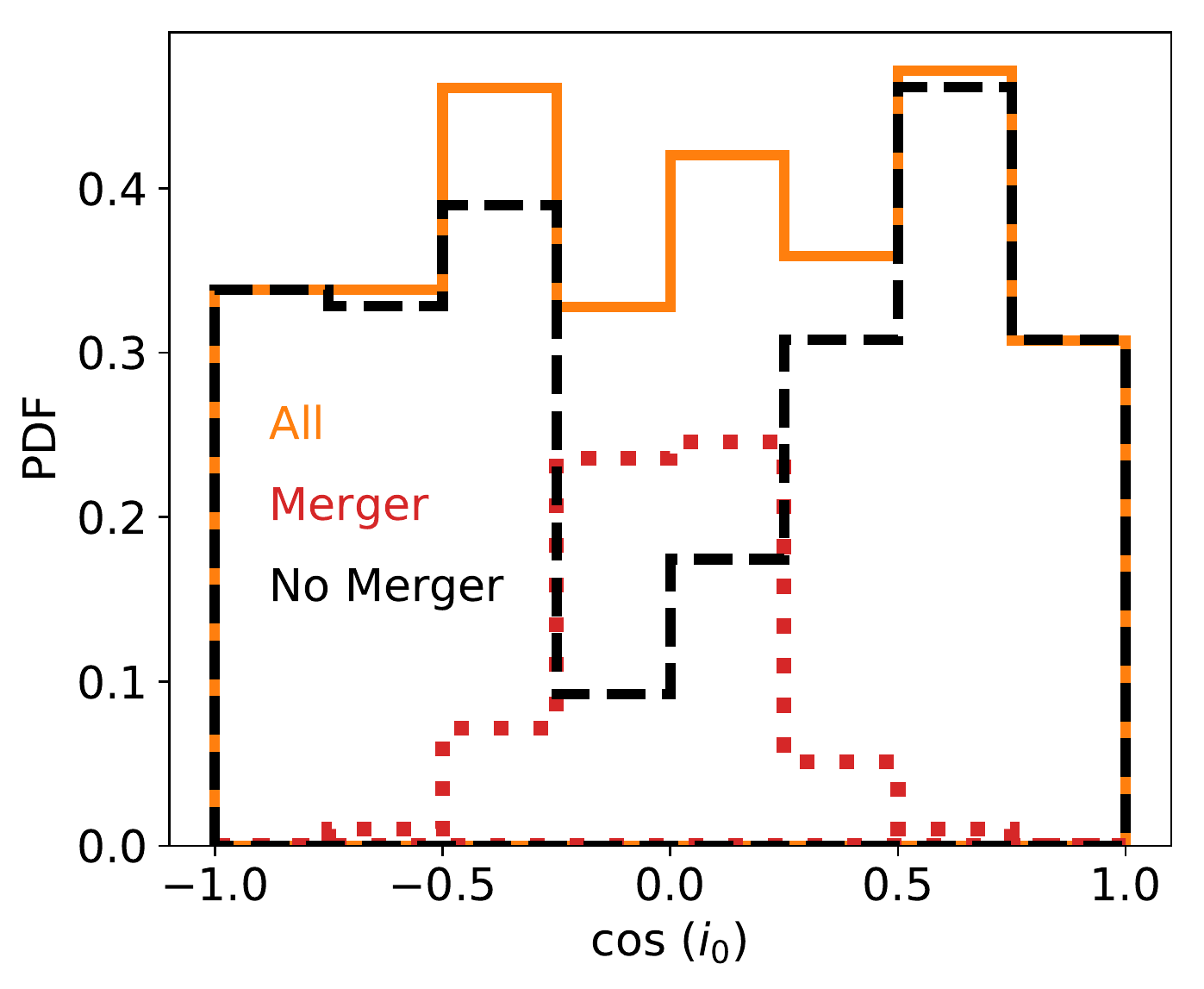}
\caption{Top panel: initial inclination distribution of massive binaries that merge in four different simulation runs. Most of the systems merge when $i_0\sim 90^\circ$ as a consequence of the excursions to very high eccentricity due to the LK cycles. Bottom panel: comparison between systems that merge and systems that do not merge in our fiducial model.}
\label{fig:incl}
\end{figure}

In Figure~\ref{fig:incl}, we show the inclination probability distribution function (PDF) of merging binaries orbiting an SMBH of mass $\msmbh=4 \times 10^6 \msun$ (Milky Way-like galaxy), for four different models. We label each model with the typical parameter that discriminates it with respect to our fiducial model. As expected, the majority of the mergers take place when the initial relative inclination is $\sim 90^\circ$, almost independently of the details of the models. Under this circumstance, the LK effect is the maximum, the eccentricity oscillates up to unity and the binaries merge near the pericentre. When considering the octupole order of approximation, some systems can undergo strong excursions in the inner eccentricity, and thus merge, if the outer orbit is eccentric, even if the initial inclination is outside of the mentioned window \citep{naoz13a}. \citet{steph2016M} found that systems that merge have typically large values of the octupole parameter, of the order $\epsilon\sim 10^{-3}$. For the binaries that merge in our simulations, we have  $10^{-4}\lesssim\epsilon\lesssim 10^{-3}$, (note that the lower limit must be 0) and 
contrary to \citet{steph2016M}  we do not find evidence for a  dependence
of the merger numbers on $\epsilon$.  In order to see the effect of the octupole terms, in Figure~\ref{fig:incl} we compare the inclination distribution of the equal mass case (i.e., $m_1=m_2$) to the one obtained by assuming a uniform mass ratio distribution. In the former case, $\epsilon=0$ and the octupole contribution is identically zero (see Eq.\ \ref{oc1}). We do not find significant differences in the inclination probability distributions in the two cases. Moreover,
the number of systems that merge in the equal mass case
is comparable to the other models (see Table 1). These results indicate therefore that the effect of the octupole order terms is not great in our simulations \citep[in agreement with][]{2017ApJ...846..146P}. 
For other choices of initial conditions
 the octupole order terms could become more important \citep{steph2016M}.
 Moreover, we are not limited to any order of approximation of the 3-body equations of motion, as in previous works, and there will be  a non-zero contribution from higher order terms (beyond the octupole) even for  $m_1=m_2$.

In the bottom panel of Figure~\ref{fig:incl}, we show the inclination probability distribution for both the systems that merge and that do not merge, and the total sampled systems, in our fiducial model. Since the majority of the mergers take place when the initial relative inclination is large, the distribution of the inclinations of the surviving massive binaries results to be nearly depleted in high-inclined systems ($\cos i_0\sim 0$).

\subsection{Mass distribution}

\begin{figure} 
\centering
\includegraphics[scale=0.55]{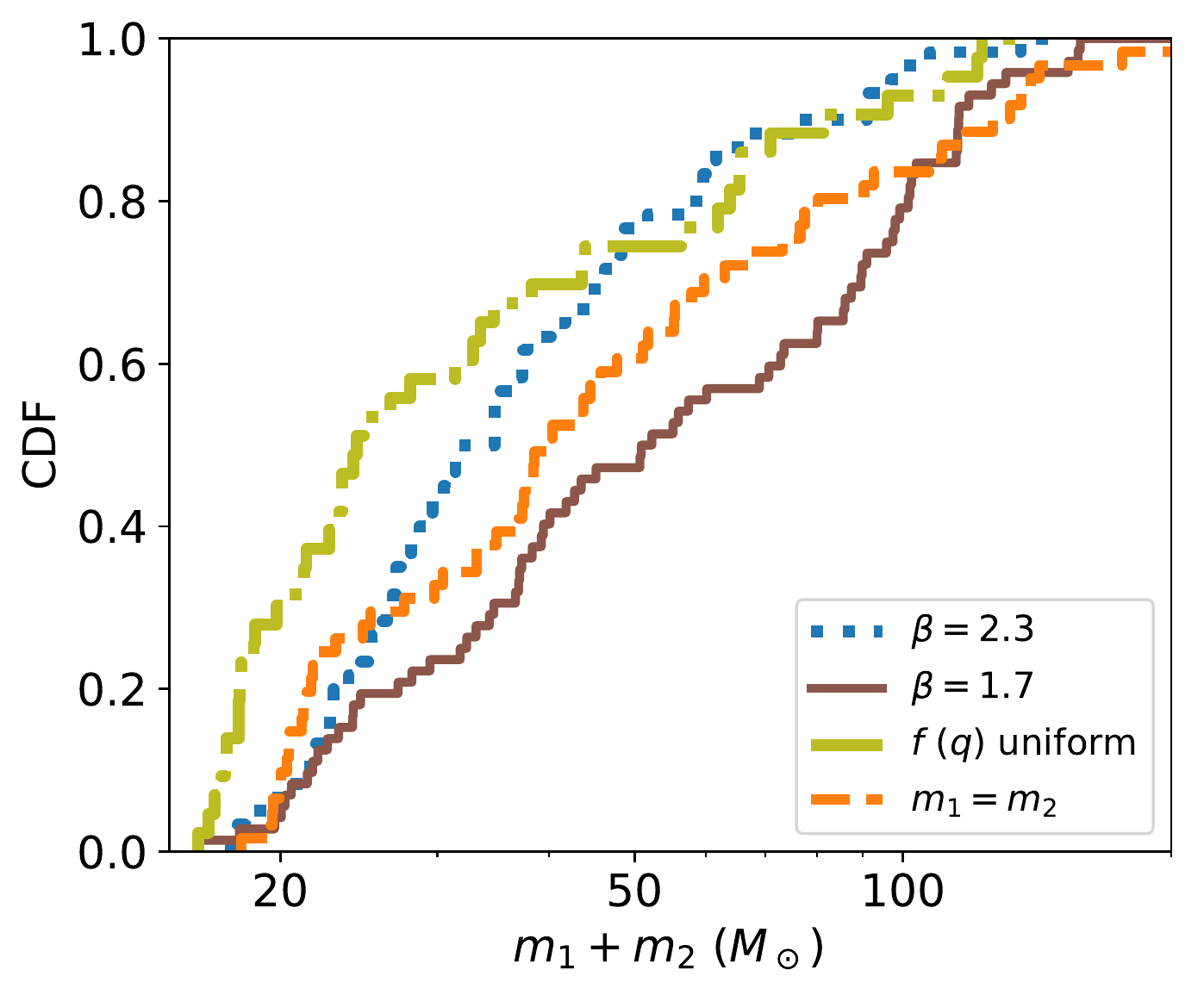}
\caption{Cumulative distribution function of the total mass ($m_1+m_2$) of merging binaries in a Milky Way-like nucleus for different slopes of the initial mass function ($\beta$) and mass-ratio distribution ($f(q)$). Note that we label each model with the typical parameter that discriminates its with respect to our fiducial model.}
\label{fig:mass}
\end{figure}

In our runs, we consider two different slopes of the initial mass function and different pairing of the stars in the binaries, as described in Section~\ref{sect:binprop}. Figure~\ref{fig:mass} illustrates the cumulative distribution function (CDF) of the total mass $m_1+m_2$ of merging binaries for a Milky Way-like nucleus for different slopes of the initial mass function ($\beta$) and mass-ratio distribution ($f(q)$). 

As expected, we find that a steeper initial mass function implies a steeper distribution of the total mass of merging binaries, since the mass distribution prefers smaller stellar masses. We found that the distributions of merging binaries nearly map the initial distributions. The latter are not affected by the stability criterion of Eq.~\ref{eqn:hills}. In our fiducial model ($\beta=2.3$ and random pairing of stars), we find that $\sim 50\%$ and $\sim 90\%$ of the mergers have total mass $\lesssim 30\msun$ and $\lesssim 70\msun$, respectively, while $\sim 50\%$ and $\sim 90\%$ of the mergers have total mass $\lesssim 70\msun$ and $\lesssim 120\msun$, respectively, when we consider a slope $\beta=1.7$.

The distribution of total mass is also affected by the mass-ratio distribution. We find that $\sim 50\%$ and $\sim 90\%$ of the mergers have total mass $\lesssim 25\msun$ and $\lesssim 70\msun$, respectively, when we sample the mass-ratio from a uniform distribution, whereas $\sim 50\%$ and $\sim 90\%$ of the mergers have total mass $\lesssim 40\msun$ and $\lesssim 125\msun$, respectively, when the stars in the binary have the same initial mass. The CDF of our fiducial model ($\beta=2.3$ and random pairing of stars) lies in between the models where the mass-ratio is drawn from a uniform distribution and where $m_1=m_2$. The former typically predicts smaller total masses, the latter larger total masses. These curves follow from the initial distributions of the stellar masses. In the case $f(q)$ is uniform, the mass of the secondary is always smaller than the primary, resulting in a smaller total binary mass than the case where the mass of the secondary is sampled independently from the initial mass function. On the other hand, the total binary mass is typically larger when $m_2=m_1$. This is true in particular for large total mass, since very massive stars are disfavoured with respect to less massive stars from the initial mass function.

\subsection{Semi-major axis and eccentricity}

\begin{figure} 
\centering
\includegraphics[scale=0.55]{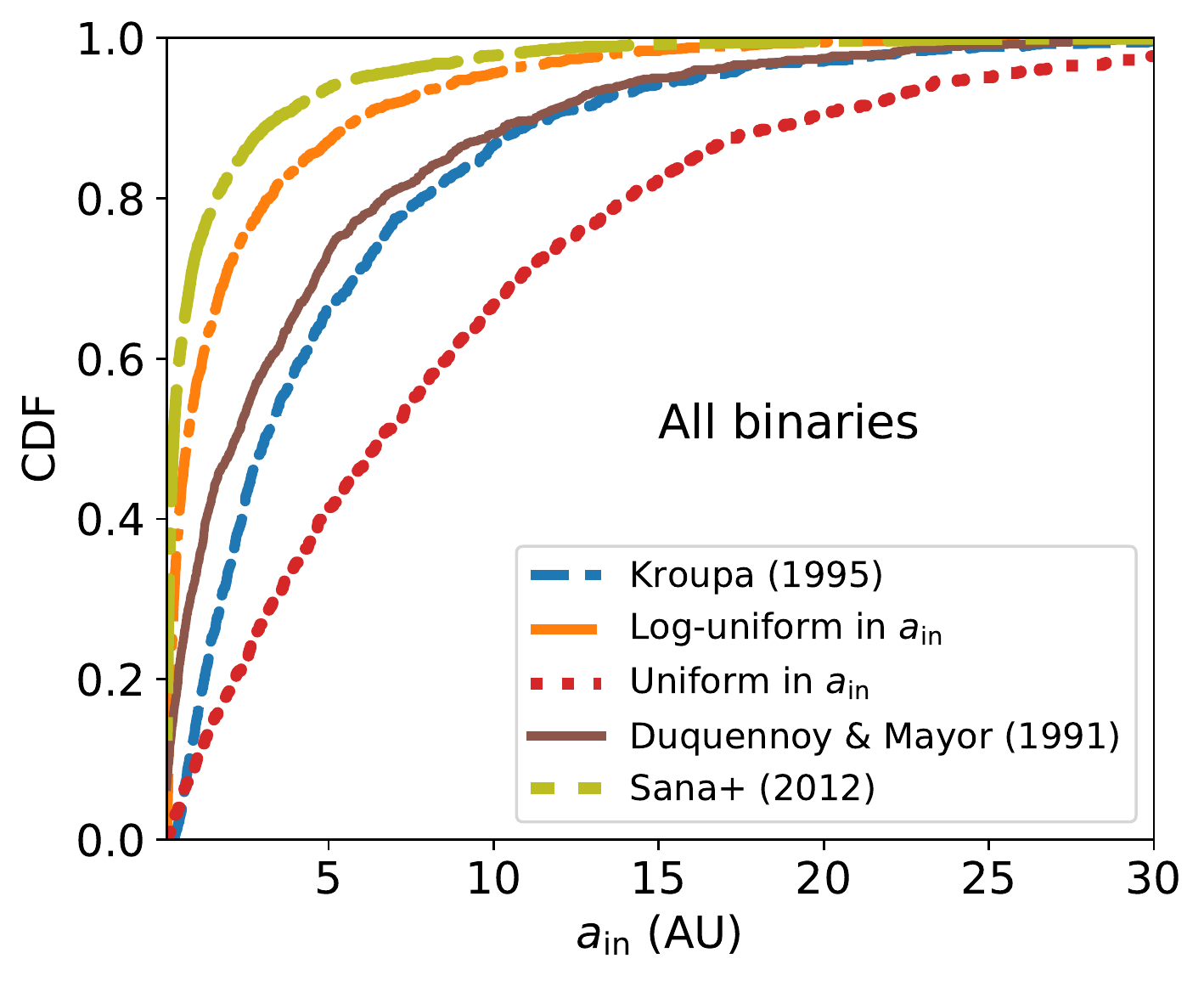}
\includegraphics[scale=0.55]{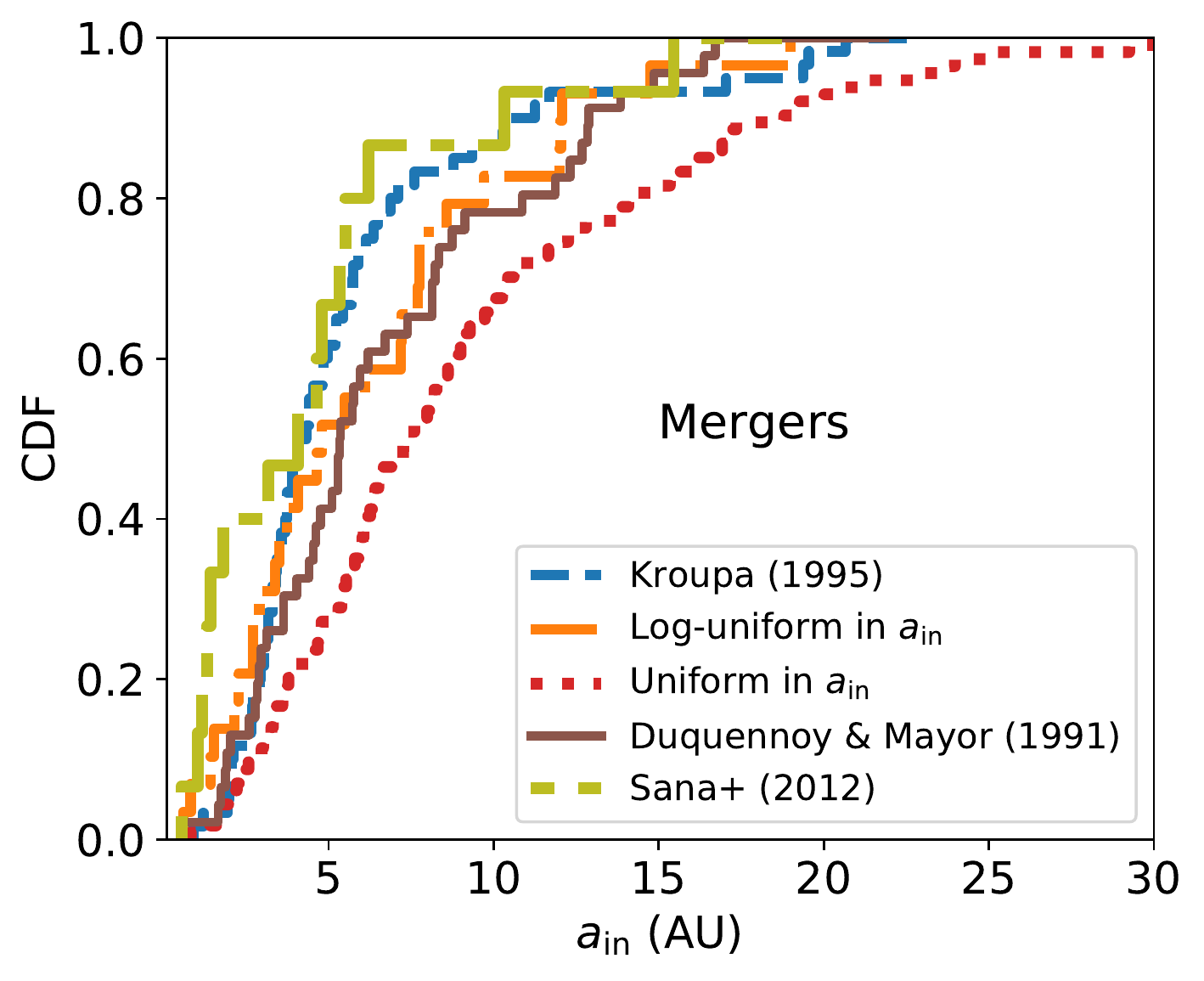}
\caption{Cumulative distribution function of the initial inner semi-major axis ($\ain$) of all binaries (top) and merging binaries (bottom) for a Milky Way-like nucleus and different initial inner semi-major axis distributions, as described in Section \ref{sect:binprop}. Note that we label each model with the typical parameter that discriminates it with respect to our fiducial model. A very few systems have $\ain>30$ AU.}
\label{fig:semin}
\end{figure}

\begin{figure} 
\centering
\includegraphics[scale=0.55]{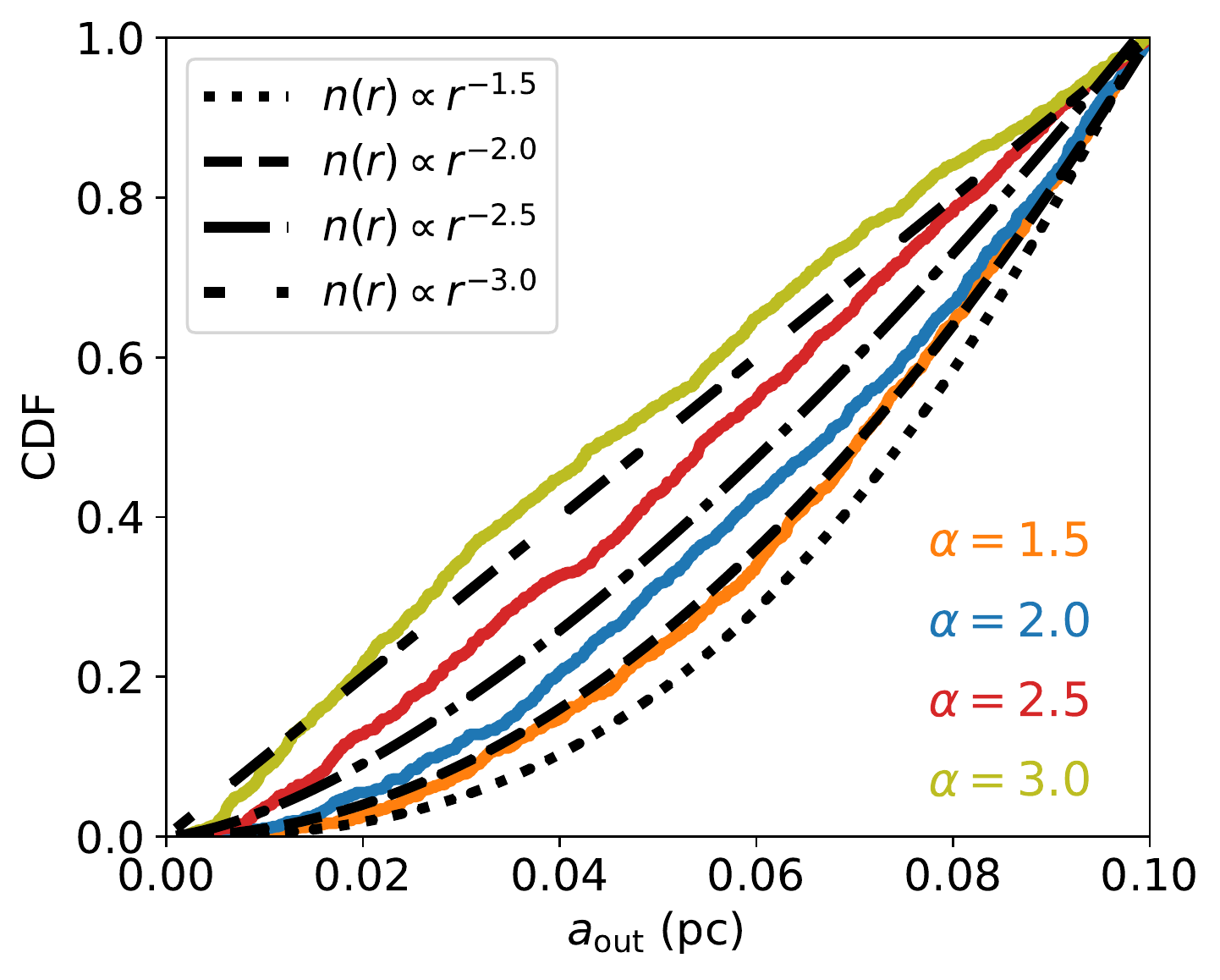}
\includegraphics[scale=0.55]{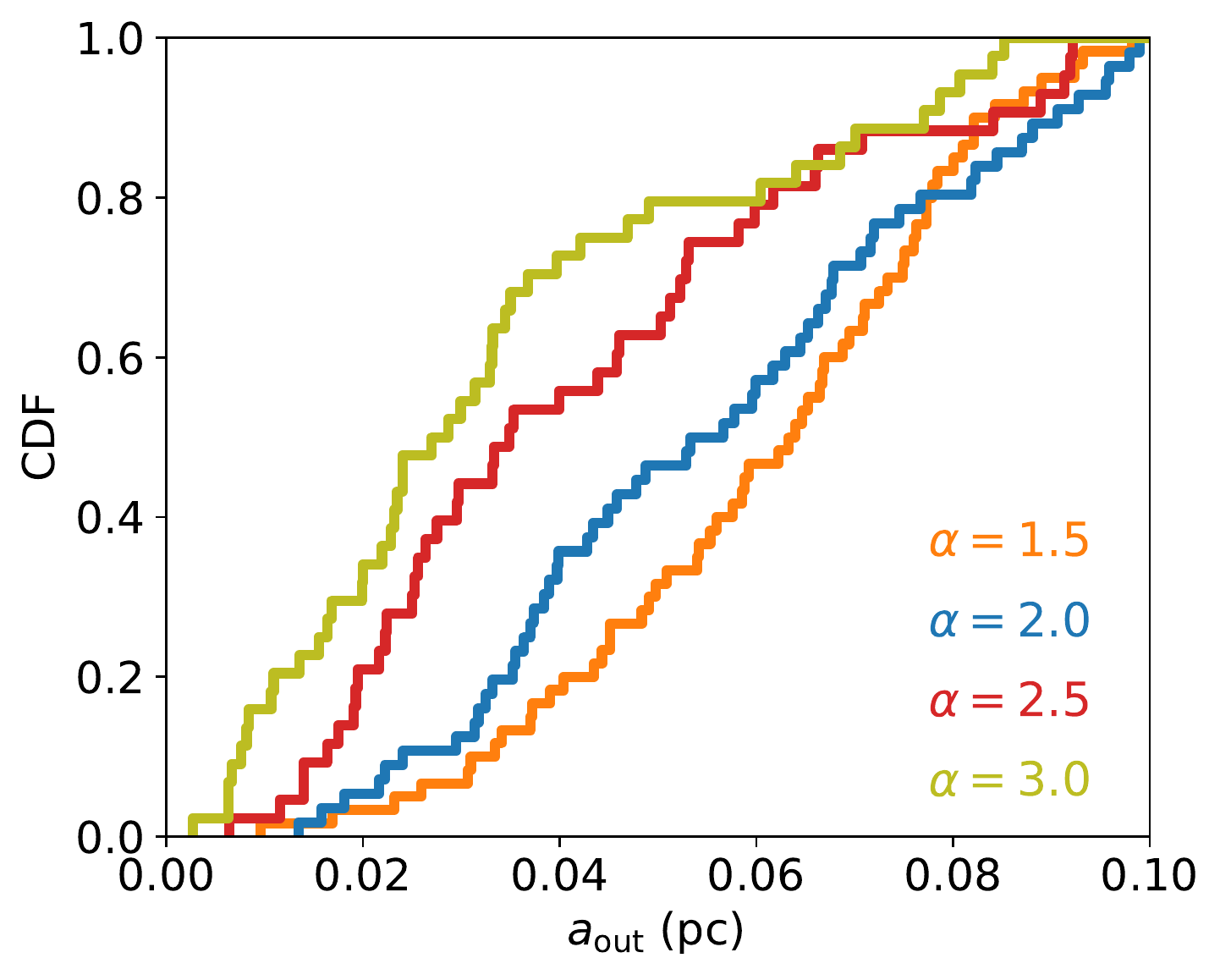}
\caption{Cumulative distribution function of the initial outer semi-major axis ($\aout$) of all binaries (top) and merging binaries (bottom) for a Milky Way-like nucleus and different values of the slope $\alpha$ of the binary density distribution around the SMBH. Note that we label each model with the typical parameter that discriminates it with respect to our fiducial model.}
\label{fig:semout}
\end{figure}

We adopt five different prescriptions for the initial distribution of the inner semi-major axis of the binaries as described in Section \ref{sect:binprop}. Figure~\ref{fig:semin} reports the CDF of the initial inner semi-major axis ($\ain$) of all binaries (top) and merging binaries (bottom) for a Milky Way-like nucleus and different initial inner semi-major axis distributions. The top panel refers to the inner semi-major axis distribution after applying the stability criterion of Eq.~\ref{eqn:hills}. This tends to favour tighter binaries than predicted by the relative nominal distributions, since tighter binaries are more stable than wider binaries for a given initial semi-major axis distribution. We found a very few systems that have $\ain\gtrsim 30$ AU. The distribution of merging binaries has a similar shape regardless of whether  $\ain$ is sampled from a \citet{kroupa1995}, \citet{duq1991}, \citet{sana2012}, or a log-uniform distribution. We find that $\sim 50\%$ of the merging binaries have $\ain\lesssim 5$ AU. For the \citet{kroupa1995} and \citet{duq1991} models, the CDFs approximately map the initial distributions of $\ain$. Also the \citet{sana2012} and the log-uniform models have a similar CDF. As shown in the top panel of Figure~\ref{fig:semin}, these models prefer tighter binaries with respect to the other models, and for a larger fraction of binaries the LK cycles are suppressed by apsidal precession due to tides and/or GR. As a result, the distribution of $\ain$'s for merging binaries are comparable to the other models, and, additionally, produce the smallest merger fractions (see Tab.~\ref{tab:models}). We also find that $\sim 50\%$ and $\sim 90\%$ of the merging binaries have $\ain\lesssim 8$ AU and $\ain\lesssim 20$ AU, respectively, when the initial semi-major axes are drawn from an uniform distribution in $\ain$. These values are larger than the reference values for the other models and reflect the fact that the initial semi-major axes are typically larger when sampled from an uniform distribution in $\ain$. 

In our fiducial model, we have assumed that the massive binaries follow a power-law density cusp around the SMBH with $\alpha=1.5$. Since the exact slope of the density cusp is not well know, we study how varying the $\alpha$ in the range $1.5$--$3$ would affect the results. The slope of the cusp affects mainly the typical distance from the SMBH where massive binaries undergo a merger event. Figure~\ref{fig:semout} shows the CDF of the initial outer semi-major axis ($\aout$) of all binaries (top) and merging binaries (bottom) for a Milky Way-like nucleus and different values of the slope $\alpha$ of the binary density distribution around the SMBH. The stability criterion (Eq.~\ref{eqn:hills}) slightly affects the nominal distributions of outer semi-major axis. The distributions of $\aout$ of merging binaries nearly maps the initial distribution. In our runs, we find that $\sim 50$\% of the mergers happen within $\sim 0.07$ pc, $\sim 0.045$ pc, $\sim 0.025$ pc, $\sim 0.02$ pc for $\alpha=1.5$, $\alpha=2.0$, $\alpha=2.5$, $\alpha=3.0$, respectively. As expected, shallower cusps predict a larger number of mergers further from the SMBH compared to steeper cusps.  The reason is that steeper cusp distributions (larger $\alpha$'s) imply smaller outer semi-major axes for the binaries. This in turn implies smaller LK timescales and more rapid oscillations in eccentricity, thus mergers. We note, however, that for every value of $\alpha$ considered in this work most of the binaries undergo at least one LK timescale within their lifetime, with the possible exception of tight binaries, where the LK cycles are typically damped by GR or tidal precessions.

\begin{figure} 
\centering
\includegraphics[scale=0.55]{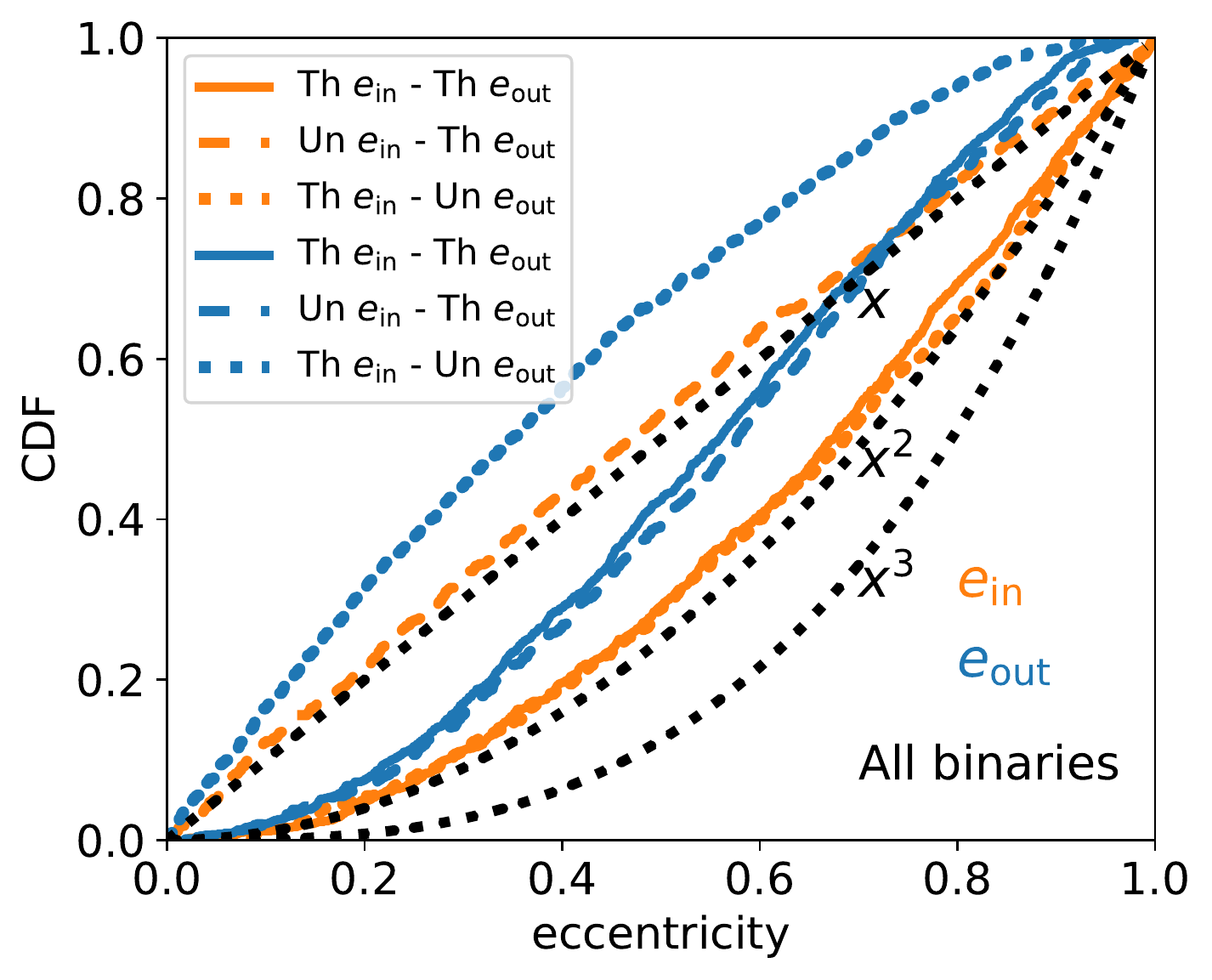}
\includegraphics[scale=0.55]{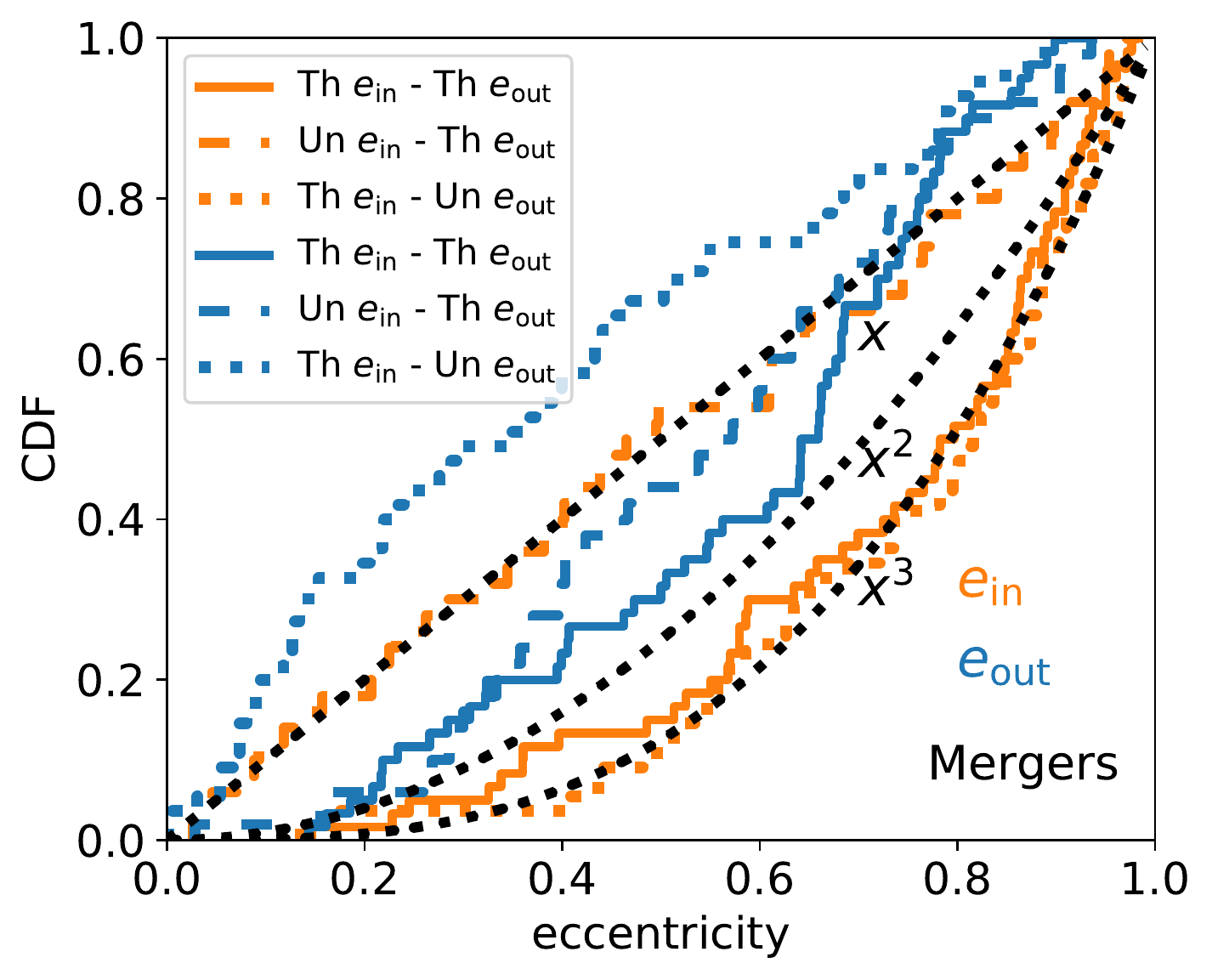}
\caption{Cumulative distribution function of the initial inner ($\ein$) and outer ($\eout$) eccentricities of all binaries (top) and merging binaries (bottom) for a Milky Way-like nucleus and different initial eccentricity distributions (Thermal: Th; Uniform: Un). Note that we label each model with the typical parameter that discriminates it with respect to our fiducial model.}
\label{fig:ecc}
\end{figure}

Inner and outer orbital eccentricities are also not well constrained in galactic nuclei. Independently of their original orbit, resonant relaxation operates on a timescale $T_{\rm RR}$ (Eq.~\ref{eqn:trr}) and the angular momentum, hence the eccentricity, of stars diffuses away from the initial value. For what concerns the inner eccentricity, it may depends on the scattering of the massive binaries by local COs and stars. As a consequence, an \textit{in-situ} formation of massive stars in a disc would be expected to favour circular binaries, while a thermal eccentricity distribution might be preferred if the bianries originated from the stellar cusp. To investigate how the results are affected by our assumptions about the initial eccentricity distribution, we run models with both a thermal and uniform distributions of eccentricities. Figure~\ref{fig:ecc} illustrates the CDF of the initial inner ($\ein$) and outer ($\eout$) eccentricities of all binaries (top) and merging binaries (bottom) for a Milky Way-like nucleus and different initial eccentricity distributions. The top panel refers to the inner semi-major axis distribution after applying the stability criterion of Eq.~\ref{eqn:hills}. We plot both  the results for a thermal (Th) distribution and an uniform (Un) distribution. The rejection criterion of Eq.~\ref{eqn:hills} does not affect the shape of the distribution of the initial inner eccentricity, both in the case it is sampled from a thermal and an uniform distribution. On the other hand, it affects the distribution of $\eout$, which prefers smaller values than the nominal case both from an uniform and thermal distribution. In the case of a thermal distribution, our results show that massive binaries that merge have typically larger initial inner (outer) eccentricities. However, while the $\eout$ distribution of the merging binaries maps the initial distribution (after Eq.~\ref{eqn:hills}), the $\ein$ distribution does so only in the case it is sampled from an uniform distribution. When a thermal distribution of the inner eccentricity is taken into account, we find a superthermal distribution ($\propto e_{\rm in}^3$) for the merging binaries.

\subsection{Merger times}

\begin{figure} 
\centering
\includegraphics[scale=0.55]{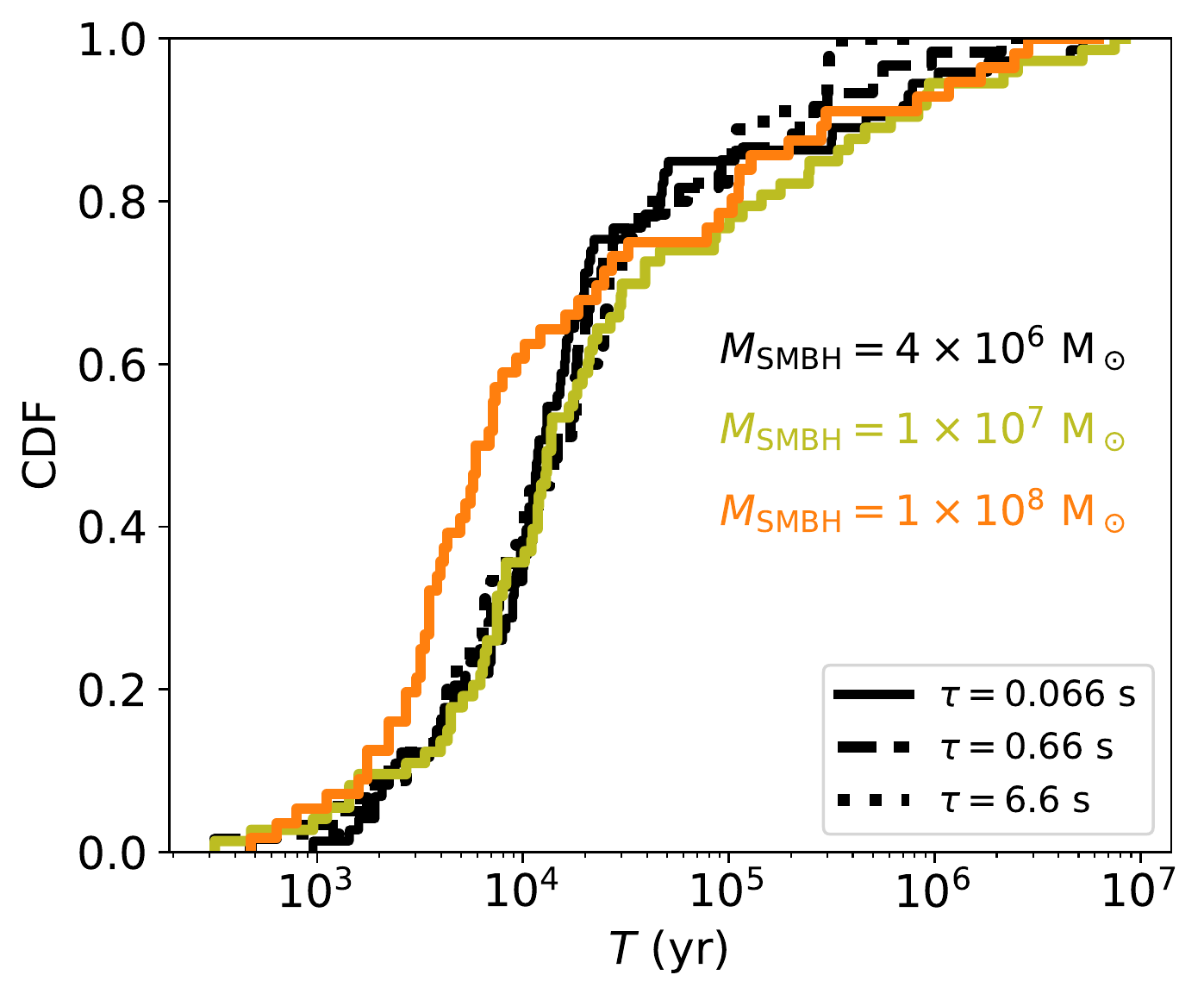}
\caption{Cumulative distribution function of merger times of merging binaries for different SMBH masses ($\msmbh$) and time-lag parameter ($\tau$). Note that we label each model with the typical parameter that discriminates its with respect to our fiducial model.}
\label{fig:time}
\end{figure}

The SMBH mass can affect the distribution of merger times, since more massive SMBHs are expected to perturb stronger the massive binaries and have a shorter LK timescale. Nevertheless, massive binaries orbiting heavier SMBHs have also a tighter stability region. From Eq.~\ref{eqn:hills}, the minimum outer semi-major axis is $\propto \msmbh^{1/3}$, which combined with Eq.~\ref{eqn:tlk} gives $T_{\rm LK}\propto P_{\rm out}^2\propto a_{\rm out}^3/\msmbh\propto 1$. As a consequence, we do not expect a very significant dependence on  $\msmbh$. We show this in Figure~\ref{fig:time}, where we plot the merger time CDFs of massive binaries that lead to a merger for different $\msmbh$'s. To further check this, we have compute Eq.~\ref{eqn:tlk} for all the systems drawn for different $\msmbh$'s and found that statistically $T_{\rm LK}$ for $\msmbh=10^8 \msun$ is typically only a factor $\sim 2$ smaller than the $T_{\rm LK}$ for $\msmbh=4\times 10^6 \msun$.

Moreover, the typical merger time can be affected by the adopted value of the time-lag $\tau$.

Figure~\ref{fig:time} also shows the merger time CDFs of massive binaries that lead to a merger as a function of $\tau$ for $\msmbh=4\times 10^6\msun$. The merger time distributions do not depend on the choice of the time-lag parameter. This can be justified by looking at the typical timescale for dissipative tides (Eq.~\ref{eqn:tfric}). For example, for a massive binary comprised of two stars of $10\msun$ each, using Eq.~\ref{eqn:tfric} we find that  $T_{\rm fric}\sim 5\times 10^{14} (0.66\ \mathrm{s}/\tau)(\ain(1-e_{\rm in}^2)/\mathrm{AU})^8$ yr. Thus, the inner semi-major axis has to be quite small in order for the dissipative process to be efficient during the MS lifetime of the massive stars ($\sim$ few Myr) and to be shorter than the LK timescale.

\subsection{Rates}

We can use the results of our $N$-body simulations combined with the observations to estimate the rate at which binaries merge due to LK oscillations within a given distance from the SMBH. Even though the star-formation rate close to non-resolved regions around SMBHs is difficult to estimate, we use an empirical estimate based on our Galactic Centre \citep[see e.g.][]{bart09}. Approximatively $\sim 200$ O-stars (likely to later form NSs and BHs) are observed and inferred to have formed over the last $\sim 10$ Myrs in the young stellar disk at distances $\sim 0.05$--$0.5$ pc from the SMBH. The number of lower-mass B-stars in the same environment suggests that similar continuous star-formation may have not occurred over the last $\sim 100$ Myr. Based on our Galactic Centre, we may consider an in-situ formation rate of massive stars of $\Gamma_{\rm sup}\sim 200/10^8=2\times10^{-6}$ yr$^{-1}$. A similar rate can be computed in the case massive stars are supplied by 2-body migration \citep[ex-situ scenario, see e.g.][]{antper2012,fragrish2018}. Hence, the overall rate of massive binary star mergers is
\begin{eqnarray}
\Gamma(\msmbh)&=&\Gamma_{\rm sup}f_{\rm mb}f_{\rm merge}=1.4\times10^{-7} \yr^{-1}\times\nonumber\\
&\times &\left(\frac{4\times 10^6\msun}{\msmbh} \right)^{1/4}\left(\frac{f_{\rm mb}}{0.7}\right)\left(\frac{f_{\rm merge}}{0.1}\right)\ ,
\label{eqn:rate}
\end{eqnarray}
where $f_{\rm mb}$ is the fraction of massive stars in binaries, that we normalize to the value observed in the field \citep{duq1991,sana2012}, and $f_{\rm merge}$ is the fraction of mergers we find in our simulations (see the last column in Tab.~\ref{tab:models}). Our results show that the fraction of merging massive binary stars is in the range $0.04$--$0.15$, regardless of the details of the initial distributions of masses and orbital elements. For a Milky Way-like nucleus, we find that the typical rate of massive binary star mergers is $\Gamma\approx 1.4\times10^{-7} \yr^{-1}$. If the merger rate is $\sim 1.4\times10^{-7} \yr^{-1}$, then there is one stellar merger produced every $\sim 7$ Myr. The age of the S-star cluster is uncertain but in the range $10$--$100$ Myr \citep{gill2017}, and therefore $\sim 1$--$10$ currently observed S-stars could be the product of mergers.

\section{Discussion}
\label{sect:disc}

A number of authors have studied the products and the fate of the collisions in galactic nuclei. \citet{2010ApJ...713...90A} first studied the evolution of stellar binaries orbiting an SMBH in a nuclear stellar cluster using a method similar to the one adopted in this paper but without 
the tidal terms and for a limited set of initial conditions. \citet{prod2015}, mainly using secular equation of motions at the quadrupole order,  found that stellar binaries in this environment can significantly change their orbital evolution due to the SMBH acting as a distant perturber, and can efficiently merge when their orbital plane is highly inclined with respect to orbital plane of the binary around the SMBH, consistent with the predictions of the LK theory. \citet{steph2016M} later updated this study by using the secular equations at the octupole level of approximation and showed that LK oscillations lead $\sim 13\%$ of system to merge on a timescale of $\sim 6$ Myr. They also considered the post-main-sequence evolution of tidally locked binaries, using standard prescriptions for the binary stellar evolution \citep{hurl2002}. These binaries may undergo a collision when one of the stars overflows its Roche lobe, which increases the merger fraction to $\sim 18\%$ on a timescale of $\sim$ few Gyr. The fate of merging binaries has recently been further investigated by \citet{steph2019}.

In this paper, we have run full high-precision $N$-body simulations of the 3-body system consisting of a binary star and the SMBH.  This allowed us to precisely follow the high-eccentricity excursions of the binary due to LK cycles, which often occur in the portion of the phase space where the secular equations of motion are not a precise description of the 3-body dynamics \citep{2014ApJ...781...45A,antognini14,fragrish2018}. While previous similar work considered only one or a couple of models \citep[e.g.,][]{2010ApJ...713...90A}, we have investigated $17$ different models. Here, we have explored the role of the different assumptions on the massive binary properties, which are still uncertain to some extent, and the role of the mass of the perturbing SMBH. We have found a merger fraction in the range $0.04$--$0.15$ on a timescale of $\sim 7$ Myr, consistent with previous work. We have also showed how the different initial binary properties affect the typical mass, inner and outer semi-major axes, and inner and outer eccentricities of merging systems.

\citet{2011ApJ...731..128A} used SPH simulations to follow the collision of stars due to the LK mechanism near a SMBH. In mergers involving a low-mass and a high-mass star, the merger product was found to acquire a high core hydrogen abundance from the smaller star, effectively resetting the nuclear evolution "clock" to the zero age main-sequence. The products of binary mergers can become therefore  a rejuvenated star, more massive than each of their original progenitors, and could possibly be observed as blue stragglers \citep{perfab2009}. Moreover, the formation of young massive merger products can affect the observed stellar mass-function of the O and B stellar populations and render it to be top-heavy. This could definitively be important to constrain the dynamics and star formation history of galactic nuclei. In our Galactic Centre, \citet{lu2013} found evidence of a young clusters of massive stars, of age $2.5$ Myr--$5.8$ Myr, with an initial mass function with slope $1.7\pm 0.2$, much steeper than the canonical \citet{kroupa2001} mass function. Interestingly, also the observed G2 in our Galactic Centre could be explained by binary star mergers. G2 is an extremely red object travelling on a very eccentric orbit ($e\sim 0.98$) of semi-major axis $\sim 0.03$ pc \citep{gill2012,gill2013}. As suggested in \citet{phi2013} and discussed in \citet{prod2015}, G2 could be the result of the merger of two stars, whose high-eccentric orbit around the SMBH could be explained with an initial tidal separation of a triple or quadruple star \citep{perets2009,fff2018,fgu18}.

Observing the merger products of binary stars in galactic nuclei could not be easy. Some Local Group galactic nuclei have been observed to harbour enigmatic stellar populations of blue stars. \citet{lassm16} considered the origins of enigmatic stellar populations in four Local Group galactic nuclei (Milky Way, M31, M32 and M33), where in three out of the four nuclear star clusters centrally concentrated blue stars have been observed. \citet{lassm16} calculated order-of-magnitude estimates for various collision rates, as a function of the host environment and distance from the centre of the host nucleus. They showed that the rates are sufficiently high that blue stragglers formed via collisions between MS stars could contribute non-negligibly ($\sim 1$--$10$ per cent in mass) to every surface brightness profile, with the exception of the Milky Way. From Eq.~\ref{eqn:rate}, we infer that $\sim 1400$ blue straggles have been produced in our Galactic Centre over a Hubble time. For Andromeda, which hosts an SMBH of $\sim 1.4\times 10^8\msun$ in it centre, this number reduces to $\sim 580$ over a Hubble time. Blue stragglers produced though LK oscillations of massive binaries could be a significant fraction of the total number of young stars in the nuclear cluster. As a consequence, they can contaminate the population of young massive stars and should be taken into account when considering stellar age estimates and gradients. We note that also star-star encounters close to an SMBH could produce blue stragglers \citep{alex17}. In both cases, the final mass of the merger product depends on the relative mass ratio of the colliding objects and on the ratio between their relative velocity to their surface escape speed \citep{benz1987,trac2007,gabu2010}. These collisions have been shown to possibly play some role also in the depletion of the red giant population \citep{dale2006,bart2010}. Unfortunately, given the complex stellar population of galactic nuclei, where there could be continuous star formation, detecting blue stragglers could be very challenging compared to the case they reside in globular clusters, where they are usually observed. Here, they presence is inferred by looking at the cluster colour-magnitude diagram, where they appear brighter and bluer than the main sequence turn-off.

Massive binaries that survive the LK cycles within their lifetime could be observed as binary S-stars, whose likely masses are of the order of $\sim 10\msun$ \citep{ginsb2016}. Inside the central arcseconds of our Galactic Centre, a group of $\sim 40$ S-stars have been observed to move on eccentric and randomly oriented orbits around the SMBH. The innermost star, the so-called S2, is a bright B-type star on an orbit of period $\sim 16$ yr and eccentricity $\sim 0.88$. \citet{gill2017} has recently provided updated measurements of the whole S-cluster. The origin of S-stars is still unknown and highly debated \citep[see e.g.][]{peretal2009}, as they could have formed in-situ, as a result of stellar formation near an SMBH, or ex-situ, e.g. from binary disruptions \citep{frasari18}. Some binaries may lurk among known S-stars. S0-27 has recently been suggested to be a candidate binary \citep{jia2019}. Binaries could also help explaining the properties of the disk of young stars observed in the GC \citep{naoz2018}. If these binaries actually exist, our model predicts them to have rather circular orbits since tidal forces between the binary companions tend to shrink and circularize their orbits, if their orbit is eccentric or their eccentricity is pumped up by moderate LK cycles. We also expect that their semi-major axis distribution would be peaked towards smaller semi-major axis because of the effects of the tides that tend to shrink the binary orbit, in particular when excited to high eccentricities by LK cycles. Furthermore, as shown in Fig.~\ref{fig:incl}, the distribution of relative inclination of the binary orbital plane would be lacking of high-inclined systems, since the majority of the mergers take place when the initial relative inclination is large. The fraction of these systems depends on the typical star structure and composition, parametrized by $\tau$ (or $Q_{\rm tide}$), and a statistical sample of them can be used to put constraints on the average stellar intrinsic composition in binaries.

On long timescales, the post-MS evolution of these binaries affects their properties \citep{steph2016M}. Some of them will explode as a supernova and can produce a NS or a BH, if the mass if high enough. If the SN natal velocity kicks are not too high, the system can survive as a binary of COs, otherwise it becomes unbound \citep{fragl2019,lu2019}. If they stay bound, the dynamical effects due to the non-coherent scattering with other stars and COs become important on resonant and 2-body timescales, which reshuffle the orbital paramters of the binary (Eqs.~\ref{eqn:t2b}--\ref{eqn:tvrr}). Some of the newborn unbound COs can ejected from the nuclear star cluster if the natal kick is strong enough, while other can interact with other stars and COs and possibly form new binaries. If the CO acquires a stellar companion, it can shine as an X-ray source, in particular if LK cycles are active and produce strongly interacting binaries \citep{muno2005}. The recent discovery of a cusp of X-ray binaries in the centre of our Galaxy by \citep{hailey2018}, nicely explained by \citet{ger2018}, has put attention back to these kinds of sources. If the CO binary (BH-BH, NS-NS and BH-NS) survive, the secular perturbations by the SMBH can induce LK cycles and make them merge via GW radiation emission in less than a Hubble time. The typical rate of these events has been found to be in the range $\sim 0.1$--$1$ Gpc$^{-3}$ yr $^{-1}$ \citep{antper2012,fragrish2018,hoa18}. We have shown that $f_{\rm merge}\approx 0.04$--$0.15$ of the massive binaries that lead possibly to the formation of a binary CO merge during their MS lifetime. As a consequence, the typical rate estimate for GW mergers of CO binaries in the proximity of an SMBH should be smaller by a factor $(1-f_{\rm merge})$.

\section{Conclusions}
\label{sect:conc}

We have revisited the SMBH-induced mergers of stellar binaries in galactic nuclei by means of $N$-body simulations, including tidal forces and PN terms, focusing on the massive binaries which lead to the formation of COs, that later can become a source of GWs. We have considered different SMBH masses, slopes of the IMF, distributions of the binary orbital parameters, spatial distributions of the binaries, and efficiencies in the energy dissipation due to dissipative tides. We have shown that the majority of the mergers take place when the initial relative inclination of the inner and outer orbital planes is $\sim 90^\circ$, almost independently of the details of the models, where the LK effect is the maximum, the eccentricity oscillates up to unity and the binaries merge near the pericentre. We have also found that the slope of the IMF and the mass-ratio distribution set the typical mass of merging binaries, while their orbital parameters are essentially determined by the initial distribution of inner and outer orbital elements. An interesting result is that the fraction of merging binary stars (see the last column in Tab.~\ref{tab:models}) is in the range $\sim 4\%$--$15\%$, regardless of the details of the initial distributions of masses and orbital elements. 

The origin and evolution of binaries in galactic nuclei has been under scrutiny in recent years, but it is still far from having a detailed self-consistent model that takes into account all the complex dynamics typical of the environment surrounding an SMBH. Nevertheless, mergers of binary stars and the relative products have a lot of interesting implications, in particular in the case they lead to the formation of a binary of COs, which can merge emitting GWs. Future instruments as the Keck telescope, Gemini, and the Very Large Telescope could constrain the properties of binary stars in galactic nuclei, while GW instruments, as LIGO/Virgo, will help to shed light on the BH and NS population surrounding SMBHs.

\section*{Acknowledgements}

GF is supported by the Foreign Postdoctoral Fellowship Program of the Israel Academy of Sciences and Humanities. GF also acknowledges support from an Arskin postdoctoral fellowship at the Hebrew University of Jerusalem.
FA acknowledges support from a Rutherford fellowship 
 (ST/P00492X/1) from the Science and Technology Facilities Council.
GF and FA thank Seppo Mikkola  for helpful discussions on the use of the code \textsc{archain}. We thank the anonymous referee for stimulating discussions and insightful comments. Simulations were run on the \textit{Astric} cluster at the Hebrew University of Jerusalem.

\bibliographystyle{mn2e}
\bibliography{refs}

\end{document}